\documentclass[aip,jap,graphicx,floatfix,reprint]{revtex4-1}

\usepackage{color}
\usepackage{graphicx}
\usepackage{epstopdf}
\usepackage{amsmath}

\newcommand{\appname}{$\kappa$ALD$o$}

\newcommand{\WmK}{Wm$^{-1}$K$^{-1}$} 

\begin{document}


\title{Efficient Anharmonic Lattice Dynamics Calculations of Thermal Transport in Crystalline and Disordered Solids}

\author{Giuseppe Barbalinardo}
\email{gbarbalinardo@ucdavis.edu}
\affiliation{Department of Chemistry, University of California, Davis, Davis, CA 95616, USA}%

\author{Zekun Chen}%
\affiliation{Department of Chemistry, University of California, Davis, Davis, CA 95616, USA}%

\author{Nicholas W. Lundgren}%
\affiliation{Department of Chemistry, University of California, Davis, Davis, CA 95616, USA}%

\author{Davide Donadio}
\email{ddonadio@ucdavis.edu}
\affiliation{Department of Chemistry, University of California, Davis, Davis, CA 95616, USA}%

\date{\today}

\begin{abstract}

Understanding heat transport in semiconductors and insulators is of fundamental importance because of its technological impact in electronics and renewable energy harvesting and conversion. 
Anharmonic Lattice Dynamics provides a powerful framework for the description of heat transport at the nanoscale. One of the advantages of this method is that it naturally includes quantum effects due to atoms vibrations, which are needed to compute the thermal properties of semiconductors widely used in nanotechnology, like silicon and carbon, even at room temperature.
While the heat transport picture substantially differs between amorphous and crystalline semiconductors from a microscopic standpoint, a unified approach to simulate both crystals and glasses has been devised.
Here we introduce a unified workflow, which implements both the Boltzmann Transport equation (BTE) and the Quasi Harmonic Green-Kubo (QHGK) methods. We discuss how the theory can be optimized to exploit modern parallel architectures, and how it is implemented in \appname:  a versatile and scalable open-source software to compute phonon transport in solids.
This approach is applied to crystalline and partially disordered silicon-based systems, including bulk silicon and clathrates, and on silicon-germanium alloy clathrates with largely reduced thermal conductivity. 
\end{abstract}

\maketitle

\section{Introduction}
The theory of lattice thermal transport in solids has been developed over several decades. Yet, computing heat transport in complex nanostructured materials is still challenging, as one needs to model mesoscopic systems with low or no symmetry at full atomistic resolution.   
This challenge can be successfully tackled using Molecular Dynamics (MD) simulations, which can be carried out either by first principles or with empirical potentials.\cite{Alder:1970fg, Ladd1986, Marcolongo:2015dn} 

In spite of its many advantages, such as versatility, efficient size scaling and inclusion of all orders of anharmonicity, classical MD is accurate only above the Debye temperature of the materials considered, a limitation that excludes a large set of important semiconductors at room temperature.

At temperatures well below melting or other phase transitions, vibrational properties and thermal transport in solids can be computed by anharmonic lattice dynamics (ALD). Since atomic displacements from equilibrium are much smaller than interatomic distances, they can be represented in the basis of normal modes in the harmonic approximation, using the phonons quasi-particle picture, while anharmonicity is treated as a perturbation.

Anharmonicity is responsible for temperature-dependent shifts in phonon frequencies and for line-broadening, i.e. finite phonon lifetimes, thus determining the value of the thermal conductivity of a material. 
The latter can be calculated by solving the linearized phonon Boltzmann Transport equation (BTE).\cite{Peierls:1929jv,ziman2001electrons,McGaughey:2019fa}
{The linearized BTE can be solved considering the population of each single mode out of equilibrium  interacting with a bath of modes at equilibrium, in the so-called relaxation time approximation (RTA).}
In recent years, however, several approaches have been introduced to solve the BTE beyond RTA, including the self-consistent iterative approach,\cite{Omini:1996ti, Ward:2009iw,Laurent-phph-2011, Li:2014fc} a variational method,\cite{Fugallo:jl} and a direct diagonalization approach that leads to the theory of {\it relaxons}.\cite{Cepellotti:2016bk}
A few software packages have been released that implement phonon BTE for crystals in various fashion: ShengBTE,\cite{Li:2014fc} PhonTS,\cite{phonts} Phono3py,\cite{phono3py} AlmaBTE,\cite{almabte} and Alamode.\cite{Tadano:2014cy}

Alongside with these theoretical and algorithmic developments, the popularity of the ALD-BTE approach has grown, as it enables the calculation of the thermal conductivity of solids and nanostructures by first principles\cite{Broido:2007iu, Lindsay:2019fs} and the discovery of new materials with unexpected thermal transport properties.\cite{Lindsay:2013fw} The use of first-principles ALD-BTE has been further extended to 2D nanostructures,\cite{Fugallo:2014bt,Cepellotti:2015ke,Jain:2015by,Zeraati:2016in,BinOuyang:2018ha} and to more complex systems, such as intercalated layered materials,\cite{Chen:2019dg, Sood:2019bl} and quasi-one-dimensional van der Waals systems.\cite{Ott:2019bg}

The BTE formalism, however, cannot be applied straightforwardly to amorphous and partially disordered systems, due to their lack of periodicity, for which the quasi-particle picture of heat carriers breaks down. This problem was addressed by Allen and Feldman, who proposed a successful model of heat transport in glasses based on the harmonic approximation of the heat flux operator.\cite{Allen:1993gt}
Two recent works devised unified theories that explicitly include anharmonic line broadening and generalize the Boltzmann transport approach to non-crystalline systems.\cite{Isaeva:2019dm, Simoncelli:2019kg}

In this work, we provide a unified framework to calculate thermal transport in crystals, glasses, and nanostructured solids, which takes advantage of both the BTE as well as the quasi-harmonic Green-Kubo (QHGK) method.\cite{Isaeva:2019dm} We present an efficient implementation of this unified theory, highlighting the methodological steps, and how to reduce the computational \textcolor{black}{workload} in modern computer architectures.
We introduce \appname, an open-source software able to perform scalable ALD simulations. \appname\ computes frequencies, lifetimes, generalized group velocities, heat capacities, and the thermal conductivity, for crystals and glasses, offering a phonon-resolved description, which is often fundamental to understand the underlying physical picture of thermal transport.
\appname\ implements different solvers of the BTE -- specifically, the relaxation time approximation, the self-consistent solution and the direct inversion of the scattering matrix -- and extends the capabilities of the BTE solver beyond the current limit of crystalline solids with few tens of atoms per cell. 
Furthermore, the QHGK implementation in \appname\ enables the computation of thermal transport for glasses and partially disordered materials up to several thousands of atoms. 
These capabilities are enabled by a versatile architecture that takes advantage of both CPU multithreading and GPU.
Its modular design enables \appname\ to couple seamlessly to a variety of force calculators, using either empirical potentials or density functional theory (DFT). 
Finally, we exploit \appname\ to investigate the effect of germanium substitution on the thermal conductivity of pristine and germanium-doped silicon clathrates, treated with empirical potentials. We discuss the origin of different contributions to thermal transport in these systems using the ALD approach, \textcolor{black}{ highlighting the important role of symmetry breaking in these systems. }We probe the conformity between the BTE and QHGK approaches to describe heat transport in crystalline materials and random alloys.  

In section~\ref{sec:methods} we introduce the theoretical ALD framework. Starting from the expansion of the interatomic potential around the equilibrium configuration, we derive the relevant thermal quantities. We discuss the solution of the BTE using different methods and we then extend the framework to include the treatment of amorphous and partially disordered systems, using QHGK.
The implementation of the theory into an open-source package, \appname, is briefly outlined in section~\ref{seq:app}, where the main features of this project are described. 
In section~\ref{sec:application}, we apply the aforementioned methods to three classes of systems, with very different thermal properties: the silicon diamond, the pristine Type-II Silicon clathrate, and different configurations of Si-Ge Type-II clathrate random-alloy.

\section{Theory}\label{sec:methods}
\subsection{Lattice Dynamics}

In semiconductors, electronic and vibrational dynamics \textcolor{black}{often occur over different time scales}, and can thus be decoupled using the Born Oppenheimer approximation. Under this assumption, the potential $\phi$ of a system made of $N_{atoms}$ atoms, is a function of all the $x_{i\alpha}$ atomic positions, where $i$ and $\alpha$ refer to the atomic and Cartesian indices, respectively. Near thermal equilibrium, the potential energy can be Taylor expanded in the atomic displacements, $\mathbf{u}=\mathbf x-\mathbf{x}_{\rm equilibrium}$, 
\begin{eqnarray}\nonumber
\phi(\{x_{i\alpha}\})&&=\phi_0 +
\sum_{i\alpha}\phi^{\prime}_{i\alpha }u_{i\alpha} 
+\frac{1}{2}
\sum_{i\alpha i'\alpha'}
\phi^{\prime\prime}_{i\alpha i'\alpha '}u_{i\alpha} u_{i'\alpha'}\\
&&+\frac{1}{3!}\sum_{i\alpha i'\alpha 'i''\alpha ''}
\phi^{\prime\prime\prime}_{i\alpha i'\alpha 'i''\alpha ''} u_{i\alpha }u_{i'\alpha '} u_{i''\alpha ''}  + \dots,\label{eq:potential_expansion}
\end{eqnarray}

where 
\begin{equation}\label{eq:second_and_third_ifc}
    \phi^{\prime\prime}_{i\alpha i'\alpha '}=\frac{\partial^{2} \phi}{\partial u_{i\alpha } \partial u_{i'\alpha '} },\qquad
    \phi^{\prime\prime\prime}_{i\alpha i'\alpha 'i''\alpha ''}=\frac{\partial^{3} \phi}{\partial u_{i\alpha } \partial u_{i'\alpha '} \partial u_{i''\alpha ''}},
\end{equation}  
are the second and third order interatomic force constants (IFC). The term $\phi_0$ can be discarded, and the forces $F = - \phi^{\prime}$ are zero at equilibrium.

The IFCs \textcolor{black}{ can be evaluated by finite difference, which consists of calculating the difference between the forces acting on the system when one of the atoms is displaced by a small finite shift along a Cartesian direction}. The second and third order IFCs need respectively, $2N_{atoms}$, and $4N_{atoms}^2$ forces calculations. In crystals, this amount can be reduced by exploiting the spatial symmetries of the system,\cite{Li:2014fc} or adopting a compressed sensing approach.\cite{Eriksson:2019hy} In the framework of DFT, it is also possible and often convenient to compute IFCs using perturbation theory.\cite{Baroni:2001tn, Paulatto:2013fb}

The dynamical matrix is the second order IFC rescaled by the masses, $D_{i\alpha i'\alpha}=\phi^{(2)}_{i\alpha i'\alpha'}/\sqrt{m_im_{i'}}$. It is diagonal in the phonons basis
\begin{equation}\label{eq:eigenvectors}
\sum_{i'\alpha'} D_{i\alpha i'\alpha'}\eta_{i'\alpha'\mu} =\eta_{i\alpha\mu} \omega_\mu^2
\end{equation}
and $\omega_\mu/(2\pi)$ is the frequency of the normal mode $\eta_\mu$ of the system. 

For crystals, where there is long-range order due to the periodicity, the dimensionality of the problem can be reduced. The Fourier transfom maps the large direct space onto a compact volume in the reciprocal space: the Brillouin zone. More precisely we adopt a supercell approach, where we calculate the dynamical matrix on $N_{\rm replicas}$ replicas of a unit cell of $N_{\rm unit}$ atoms, at positions $\mathbf R_l$, and calculate
\begin{equation}
 D_{i \alpha k i' \alpha'}=\sum_l \chi_{kl}  D_{i \alpha l i' \alpha'},\quad \chi_{kl} = \mathrm{e}^{-i \mathbf{q_k}\cdot \mathbf{R}_{l} },
\end{equation}where $\mathbf q_k$ is a grid of size $N_k$ indexed by $k$ and the eigenvalue equation becomes
\begin{equation}\label{eq:eigenvectors_crystal}
\sum_{i'\alpha'} D_{i \alpha k i' \alpha'} \eta_{i' \alpha'k s}=\omega_{k m}^{2} \eta_{i \alpha k s }.
\end{equation}
which now depends on the quasi-momentum index, $k$, and the phonons mode $s$.

\subsection{Boltzman Transport Equation}

At finite temperature $T$, the Bose Einstein statistic is the quantum distribution for atomic vibrations
\begin{equation}\label{eq:bose_einstein}
n_{\mu} = n(\omega_{\mu}) = \frac{1}{e^{{\hbar \omega_{\mu} }/{k_B  T} }- 1}
\end{equation}
where $k_B$ is the Boltzmann constant and we use $\mu =(k,s)$ to make the notation general and consistent with section~\ref{subsec:QHGK}.

We consider a small temperature gradient applied along the $\alpha$-axis of a crystalline material. If the phonons population depends on the position only through the temperature,  $\frac{\partial n_{\mu\alpha}}{\partial x_\alpha} =  \frac{\partial n_{\mu\alpha}}{\partial T}\nabla_\alpha T$, we can Taylor expand it
\begin{equation}
\tilde n_{\mu\alpha} \simeq n_\mu + \lambda_{\mu\alpha} \frac{\partial n_\mu}{\partial x_\alpha} \simeq  n_\mu + \psi_{\mu\alpha}\nabla_\alpha T
\end{equation}
with $\psi_{\mu\alpha}=\lambda_{\mu\alpha} \frac{\partial n_\mu}{\partial T}$, where $\lambda_{\mu\alpha}$ is the phonons mean free path.\cite{Srivastava:2019tl}
Being quantum quasi-particles, phonons have a well-defined group velocity, which, for the acoustic modes in the long wavelength limit, corresponds to the speed of sound in the material,
\begin{equation}\label{eq:velocity_crystal}
v_{ ks\alpha}=\frac{\partial \omega_{k s}}{\partial {q_{k\alpha}}} = \frac{1}{2\omega_{ks}}\sum_{i\beta l i'\beta'}
i R_{l \alpha} D_{i\beta li'\beta'}\chi_{kl}
\eta_{ks i\beta}\eta_{ksi'\beta}
\end{equation}
and the last equality is obtained by applying the derivative with respect to $\mathbf{q}_k$ directly to Eq.~\ref{eq:eigenvectors_crystal}.

The heat current per mode is written in terms of the phonon energy $\hbar \omega$, velocity $v$, and out-of-equilibrium phonons population, $\tilde n$:
\begin{equation}
j_{\mu\alpha'} =\sum_\alpha \hbar \omega_\mu v_{\mu\alpha'} (\tilde n_{\mu\alpha} - n_{\mu})\simeq- \sum_\alpha c_\mu v_{\mu\alpha'} \mathbf{\lambda}_{\mu\alpha}  \nabla_\alpha T .
\end{equation}
As we deal with extended systems, we can assume heat transport in the diffusive regime, and we can use Fourier's law
\begin{equation}
J_{\alpha}=-\sum_{\alpha'}\kappa_{\alpha\alpha'} \nabla_{\alpha'} T,
\end{equation}
where the heat current is the sum of the contribution from each phonon mode: $J_\alpha = 1/(N_k V)\sum_\mu j_{\mu\alpha}$.
The thermal conductivity then results:
\begin{equation}\label{eq:conductivity_from_lambda}
\kappa_{\alpha \alpha'}=\frac{1}{ V N_k} \sum_{\mu} c_\mu v_{\mu\alpha} \lambda_{\mu\alpha'},
\end{equation}
where we defined the heat capacity per mode 
\begin{equation}\label{eq:heat_capacity}
c_\mu=\hbar \omega_\mu \frac{\partial n_\mu}{\partial T},
\end{equation}
which is connected to total heat capacity through $C = \sum_\mu c_\mu /NV$.

We can now introduce the BTE, which combines the kinetic theory of gases with collective phonons vibrations:\cite{Peierls:1929jv, ziman2001electrons} 
\begin{equation}
{\mathbf{v}}_{\mu} \cdot {\boldsymbol{\nabla}} T \frac{\partial n_{\mu}}{\partial T}=\left.\frac{\partial n_{\mu}}{\partial t}\right|_{\text {scatt}}, 
\end{equation}
where the scattering term, in the linearized form is 
\begin{eqnarray}
\left.\frac{\partial n_{\mu}}{\partial t}\right|_{\text {scatt}}=&&
\\
\frac{\nabla_\alpha T}{\omega_\mu N_k}\sum_{\mu^{\prime} \mu^{\prime \prime}}^{+} &&\Gamma_{\mu \mu^{\prime}  \mu^{\prime \prime}}^{+}
\left(\omega_\mu\mathbf{\psi}_{\mu\alpha}
+\omega_{\mu^{\prime}}\mathbf{\psi}_{\mu^{\prime}\alpha}
-\omega_{\mu^{\prime \prime}} \mathbf{\psi}_{\mu^{\prime \prime}\alpha}\right) 
+\nonumber
\\
+\frac{\nabla_\alpha T}{\omega_\mu N_k}\sum_{\mu^{\prime} \mu^{\prime \prime}}^{-} &&\frac{1}{2} \Gamma_{\mu  \mu^{\prime} \mu^{\prime \prime}}^{-}
\left(\omega_\mu\mathbf{\psi}_{\mu\alpha}
-\omega_{\mu^{\prime}} \mathbf{\psi}_{\mu^{\prime}\alpha}
-\omega_{\mu^{\prime \prime}} \mathbf{\psi}_{\mu^{\prime \prime}\alpha}\right) .
\nonumber
\end{eqnarray}
 $\Gamma^{+}_{\mu\mu'\mu''}$  and $\Gamma^{-}_{\mu\mu'\mu''}$  are the scattering rates for three-phonon scattering processes, and they correspond to the events of phonons annihilation $\mu, \mu'\rightarrow\mu''$  and phonons creation $\mu \rightarrow\mu',\mu''$  
\begin{eqnarray}\label{eq:gamma_operators}
\Gamma_{\mu \mu^{\prime} \mu^{\prime \prime}}^{\pm}&&=\frac{\hbar \pi}{8} \frac{g_{\mu\mu'\mu''}^{\pm}}{\omega_{\mu} \omega_{\mu'} \omega_{\mu''}}\left|\phi_{\mu \mu^{\prime} \mu^{\prime \prime}}^{\pm}\right|^{2}, 
\end{eqnarray}
and the projection of the potentials on the phonon modes are given by
\begin{equation}\label{eq:projection_plus}
\phi^\pm
_{ksk's'k'' s''}=
\sum_{il'i'l''i''}
\frac{
\phi_{il'i'l''i''}}
{\sqrt{m_{i}m_{i'}m_{i''}}}
\eta_{i ks}\eta^{\pm}_{i'k' s'}
\eta^*_{i''k''s''}\chi^\pm_{k'l'}\chi^*_{k''l''}
\end{equation}
with $\eta^+ =\eta $, $\chi^+=\chi $ and $\eta^- =\eta^*$, $\chi^-=\chi^*$. 
The phase space volume $g^\pm_{\mu\mu^\prime\mu^{\prime\prime}}$ in Eq.~\ref{eq:gamma_operators} are defined as
\begin{eqnarray}\label{eq:phase_space}
g^+_{\mu\mu^\prime\mu^{\prime\prime}} &&= (n_{\mu'}-n_{\mu''})
\delta^+_{\mu\mu^\prime\mu^{\prime\prime}}\\
g^-_{\mu\mu^\prime\mu^{\prime\prime}} &&= (1 + n_{\mu'}+n_{\mu''})
\delta^-_{\mu\mu^\prime\mu^{\prime\prime}}, 
\end{eqnarray}
and include the $\delta$ for the conservation of the energy and momentum in three-phonons scattering processes,
\begin{equation}\label{eq:delta_deconstructed}
\delta_{ks k's' k''s''}^{\pm}=
\delta_{\mathbf q_{k}\pm\mathbf q_{k'}-\mathbf q_{k''}, \mathbf Q}
\delta\left(\omega_{ks}\pm\omega_{k's'}-\omega_{k''s''}\right),
\end{equation}
with $Q$ the lattice vectors. Finally, the normalized phase-space per mode  $g_\mu=\frac{1}{N}\sum_{\mu'\mu''}g_{\mu\mu'\mu''}$, provides useful information about the weight of a specific mode in the anharmonic scattering processes. More details about the conservation of the energy and the momentum and their numerical implementations are provided in the appendix~\ref{sec:implementation_details}.

In order to calculate the conductivity, we express the mean free path in terms of the 3-phonon scattering rates
\begin{equation}
v_{\mu\alpha} = \tilde \Gamma_{\mu\mu' }\lambda_\mu = (\delta_{\mu\mu'}\Gamma^0_\mu + \Gamma^{1}_{\mu\mu'})\lambda_{\mu\alpha},
\end{equation}
where we introduced
\begin{equation}
\Gamma^{0}_\mu=\sum_{\mu'\mu''}(\Gamma^+_{\mu\mu'\mu''}  + \Gamma^-_{\mu\mu'\mu''} ),
\end{equation}
and
\begin{equation}
\Gamma^{1}_{\mu\mu'}=
\frac{\omega_{\mu'}}{\omega_\mu}
\sum_{\mu''}(\Gamma^+_{\mu\mu'\mu''}
-\Gamma^+_{\mu\mu''\mu'}
-\Gamma^-_{\mu\mu'\mu''}
-\Gamma^-_{\mu\mu''\mu'}
).
\end{equation}In RTA, the off-diagonal terms are ignored, $\Gamma^{1}_{\mu\mu'}=0$, and the conductivity is
\begin{equation}\label{eq:RTA}
\kappa_{\alpha\alpha'} =\frac{1}{N_kV} \sum_{\mu}c_\mu v_{\mu\alpha}{\lambda_{\mu\alpha'}}
=\frac{1}{N_kV} \sum_\mu c_\mu v_{\mu\alpha} {\tau_\mu}{v_{\mu\alpha'}},
\end{equation}
where $\tau_\mu=1/2\Gamma_{\mu}^0$ corresponds \textrm{black}{to} the phonons lifetime calculated using the Fermi Golden Rule. 

It has been shown that, to correctly capture the physics of phonon transport, especially in highly conductive materials, the off diagonal terms of the scattering rates cannot be disregarded.\cite{Omini:1996ti, Ward:2009iw}
More generally, the mean free path is calculated inverting the scattering tensor
\begin{equation}\label{eq:inversemeanfreepath}
\lambda_{\mu\alpha} = \sum_{\mu'}(\tilde \Gamma_{\mu\mu' })^{-1}v_{\mu'\alpha}.
\end{equation}and the conductivity is
\begin{equation}\label{eq:full_conductivity}
\kappa_{\alpha\alpha'} =\frac{1}{N_kV} \sum_{\mu\mu'} c_\mu v_{\mu\alpha}(\tilde \Gamma_{\mu\mu' })^{-1}v_{\mu'\alpha'}.
\end{equation}
\textcolor{black}{This inversion operation requires the allocation in memory of the whole $3N_{atoms}\times 3N_{atoms}$ $\Gamma_{\mu\mu'}$ tensor}.

When the off-diagonal elements of the scattering rate matrix are smaller than the diagonal we can rewrite the mean free path obtained from the BTE as a series:
\begin{eqnarray}\nonumber
\lambda_{\mu\alpha} =&&\sum_{\mu'}\left(\delta_{\mu\mu'} + \frac{1}{\Gamma^0_\mu}\Gamma^{1}_{\mu\mu'}\right)^{-1}\frac{1}{\Gamma^0_{\mu'}}v_{\mu'\alpha} = \\=&&
\sum_{\mu'}
\left[
\sum^{\infty}_{n=0}\left(- \frac{1}{\Gamma^0_\mu}\Gamma^{1}_{\mu\mu'}\right)^n
\right]\frac{1}{\Gamma^0_{\mu'}}v_{\mu'\alpha} ,\label{eq:mfp_from_scattering_inv}
\end{eqnarray}
where in the last step we used the identity $\sum_0 q^n = (1 - q)^{-1}$.  \textcolor{black}{This series converges if and only if  $|q|=\| \Gamma^1/\Gamma^0 \|<1$, which excludes several interesting physical systems.\cite{Cepellotti:2016bk}}
Eq.~\ref{eq:mfp_from_scattering_inv} can then be written in an iterative form
\begin{equation}\label{eq:self-consistent}
\lambda^0_{\mu\alpha} = \frac{1}{\Gamma^0_\mu}v_\mu
\qquad
\lambda^{n+1}_{\mu\alpha} = - \frac{1}{\Gamma^0_\mu}\sum_{\mu'}\Gamma^{1}_{\mu\mu'} \lambda^{n}_{\mu'\alpha}.
\end{equation}
Hence, the inversion in Eq.~\ref{eq:inversemeanfreepath} is obtained by the recursive expression in Eq.~\ref{eq:self-consistent}. 
Once the mean free path is calculated, the conductivity is straightforwardly computed using Eq.~\ref{eq:conductivity_from_lambda}.
\textcolor{black}{Because of the limitations on its applicability, and its computational cost, the self-consistent method may be conveniently applied only when there is no sufficient memory to perform the full inversion of the $\Gamma$ matrix.}

\subsection{Quasi-Harmonic Green Kubo}
\label{subsec:QHGK}
In non-crystalline solids with no long range order, such as glasses, alloys, nano-crystalline, and partially disordered systems, the phonon picture is formally not well-defined. While vibrational modes are still the heat carriers, their mean-free-paths may be so short that the quasi-particle picture of heat carriers breaks down and the BTE is no longer applicable. 
In glasses heat transport is dominated by a diffusive processes in which delocalized modes with similar frequency transfer energy from one to another.\cite{Allen:1993gt} 
Whereas this mechanism is intrinsically distinct from the underlying hypothesis of the BTE approach, the two transport pictures have been recently reconciled in a unified theory, in which the thermal conductivity is written as:\cite{Isaeva:2019dm, Simoncelli:2019kg}
\begin{equation}
\label{eq:conductivity_amorphous}
\kappa_{\alpha \alpha'}=\frac{1}{V} \sum_{\mu \mu'} c_{\mu \mu'} v_{\mu \mu' \alpha} v_{\mu \mu' \alpha'} \tau_{\mu \mu'}.
\end{equation}
This expression is analogous to Eq.~\ref{eq:RTA}, where modal heat capacity, phonon group velocity and lifetimes are replaced by 
the generalized heat capacity,
\begin{equation}\label{eq:twod_heat_capacity}
c_{\mu \mu'}=\frac{\hbar \omega_{\mu} \omega_{\mu'}}{T} \frac{n_{\mu}-n_{\mu'}}{\omega_{\mu}-\omega_{\mu'}},
\end{equation}
the generalized velocities,
\begin{equation}\label{eq:generalized_velocity}
v_{\mu\mu' \alpha} = \frac{1}{{2\sqrt {\omega _\mu\omega _{\mu'}} }}\mathop {\sum}\limits_{ii'\beta' \beta'' } (x_{i\alpha } - x_{i'\alpha }) D _{i\beta i'\beta'}\eta_{\mu i\beta }\eta_{\mu'i'\beta' },
\end{equation}
and the generalized lifetime $\tau_{\mu\mu'}$.
The latter is expressed as a Lorentzian, which weighs diffusive processes between phonons with nearly-resonant frequencies:
\begin{equation}\label{eq:lorentzian_amorphous}
\tau_{\mu\mu'} = 
 \frac{\gamma_{\mu}+\gamma_{\mu'}}{\left(\omega_{\mu}-\omega_{\mu'}\right)^{2}+\left(\gamma_{\mu}+\gamma_{\mu'}\right)^{2}}
\end{equation}
where $\gamma_\mu$ is the line width of mode $\mu$ that can be computed using Fermi Golden rule. 
These equations have been derived from the Green-Kubo theory of linear response applied to thermal conductivity,\cite{Green1952, Green1954, Kubo1957a, Kubo1957b} by taking a quasi-harmonic approximation of the heat current, from which this approach is named quasi-harmonic Green-Kubo (QHGK). 
It has been proven that for crystalline materials QHGK is formally equivalent to the BTE in the relaxation time approximation and that its classical limit reproduces correctly molecular dynamics simulations both for amorphous silicon up to relatively high temperature (600 K)\cite{Isaeva:2019dm} and for nanostructured silicon membranes.\cite{Neogi:2020vz}

\textcolor{black}{
In principles, QHGK entails the same computational costs as BTE-RTA, as the main bottleneck is the calculation of phonon linewidths $\gamma_\mu$. The main advantage, however, is that QHGK can be implemented as a pure real-space approach, and one can carry out all the calculations at the $\Gamma$ point. Furthermore, for disordered systems $\kappa$ is not very sensitive to the values of $\tau_{\mu\mu'}$, so that one can scale to systems larger than 10$^4$ atoms by numerically interpolating $\gamma_\mu$ computed for smaller systems.~\cite{Isaeva:2019dm}} 

Finally, we provide a microscopic definition of the mode diffusivity,\cite{Allen:1993gt}
\begin{equation}\label{eq:diffusivity_per_mode}
D_{\mu} =\frac{1}{N_k V} \sum_{\mu'}v_{\mu\mu'} \tau_{\mu\mu'}v_{\mu\mu'},
\end{equation}
which conveniently provide a measure of the 
contribution of each mode to thermal transport.

\subsection{Classical limit}\label{sec:classical_limit}
Computing the classical limit of BTE and QHGK theories is useful to estimate quantum effects and to compare with results obtained from other classical methods, for example MD.\cite{He:2012hg}
In the classical limit all the vibrational modes have energy equal to $k_BT$, so that, instead of Eq.~\ref{eq:bose_einstein}, the phonon equilibrium populations are expresses as:  
\begin{equation}
\label{Eq:classic_n}
    n_\mu = \frac{k_BT}{\hbar\omega_\mu}.
\end{equation}
As a consequence the (generalized) modal heat capacity turns out $c_{\mu\mu'} = c_\mu=k_B$.
Furthermore, the difference in phonon populations affects the three-phonon scattering phase space (Eq.~\ref{eq:phase_space}) and, therefore, the scattering rates (Eq.~\ref{eq:gamma_operators}) and ultimately the thermal conductivity calculated using Eqs.~\ref{eq:conductivity_from_lambda} and~\ref{eq:conductivity_amorphous}.
It is worth noting that the classical expression of $\kappa$ in the QHGK theory, which can be obtained by replacing the quantum populations with Eq.~\ref{Eq:classic_n} in Eqs.~\ref{eq:twod_heat_capacity}-\ref{eq:lorentzian_amorphous}, 
was formerly derived from the Green-Kubo integral of the autocorrelation function of the heat current in the quasi-harmonic approximation.\cite{Isaeva:2019dm}

\section{Scalable and general implementation: the $\kappa$ALD$o$ package}\label{seq:app}

The theory outlined in section~\ref{sec:methods} is implemented in \appname, an efficient and scalable software package which applies the ALD framework to compute the thermal transport in crystalline and non-crystalline solids at various levels of accuracy.  \appname\ features real space QHGK calculations and three different solvers of the linearized BTE: direct inversion, self-consistent cycle, and RTA. 
We here highlight the main features of \appname, while a description of the code architecture and implementation is provided in appendix~\ref{sec:implementation_details}.
\begin{itemize}
\item {\bf Forcefields.} Using the Atomic Simulation Environment,\cite{ase-paper} \appname\ can calculate the IFCs using several {\sl ab initio} and molecular dynamics codes, thus enabling the use of first-principles DFT, empirical forcefields and semi-empirical tight-binding.\cite{Aradi:2007hu} A native LAMMPS interface is also available in the USER-PHONON package. Finally, through seamless integration with the hiPhive package, the IFC calculation can take advantage of compressing-sensing machine learning algorithms. 
\item {\bf Multithread implementation on CPU and GPU.} The algorithms are implemented using linear algebra operations on tensors, to take advantage of multithreading on GPU and CPU using Numpy, Tensorflow and optimized tensor libraries.\cite{GASmith:2018kd}
\item {\bf Memory usage.} \appname\ uses $(3N_{atoms})^2$ floating-point numbers to save the state of the system when using QHGK, $N_k^2 (3N_{atoms})^2$ for the full solution of the BTE and $N_k (3N_{atoms})^2$ when using BTE-RTA. 
\item {\bf Scaling.} The slow part of ALD simulations is the calculation of the lifetime and the scattering matrix. This step requires projecting the interatomic potential on $N^3$ phonons modes and the algorithm scales like $(N_{k}(3 N_{atoms}))^3$, because of the $3$ projections on phonons modes. In \appname\ such algorithm is implemented as $2N_k(3 N_{atoms})$ tensor multiplications of size $(N_k(3 N_{atoms}))^2$ for BTE calculations while $(3 N_{atoms})^2$ for QHGK.
\item {\bf Availability.} The code is released open-source for the community to use and contribute with edits and suggestions. It is designed on modern software best practices, and we hope to provide a development platform to implement new theory and methods. The code is available at: \href{https://github.com/nanotheorygroup/kaldo}{https://github.com/nanotheorygroup/kaldo}. 
\end{itemize}

\section{Applications}
\label{sec:application}

\subsection{Diamond Silicon}\label{sec:si_diamond}

\textcolor{black}{We compute the thermal conductivity of diamond silicon at room temperature as a first benchmark of \appname\ with first-principles DFT.}
We calculated the second order IFC using Density Functional Perturbation Theory as implemented in the Quantum-Espresso package,\cite{QE-2017} using the local density approximation (LDA) for the exchange and correlation functional and a Bachelet-Hamann-Schluter norm-conserving  pseudopotential.\cite{Bachelet:1982ht} Kohn-Sham orbitals are represented on a plane-waves basis set with a cutoff of $20$~Ry and $(8, 8, 8)$ k-points mesh. The minimized lattice parameter is $5.398$~\AA. The third-order IFC is calculated using finite difference displacement on $(5, 5, 5)$ replicas of the irreducible {\sl fcc} unit cell, including up to the 5th nearest neighbor. 
The phonon lifetimes and thermal conductivity calculations were performed with \appname, using a $(19, 19, 19)$ q-point grid. 
\begin{figure}[t!]
\includegraphics[width=0.95\linewidth]{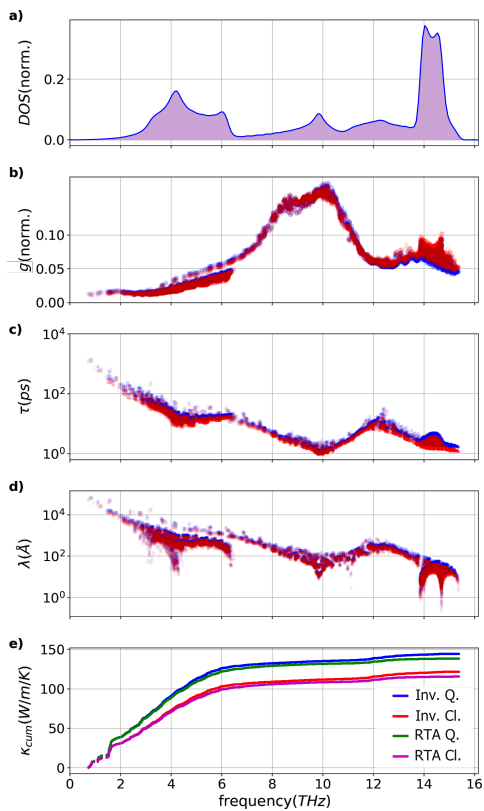}
\caption{
{Silicon diamond modes analysis. Quantum (red) and classical (blue) results are compared. a) Normalized density of states, b) Normalized phase-space per mode $g$, c) lifetime per mode $\tau$, d) mean free path $\lambda$, and e) cumulative conductivity $\kappa_{cum}$.}}
\label{fig:silicon}
\centering 
\end{figure}

The frequency-resolved phonon properties of bulk diamond Si at $300$~K are illustrated in Fig.~\ref{fig:silicon}, namely phonon density of states (a), three-phonon scattering phase-space (b), phonon lifetimes (c) and mean free paths (d). The cumulative thermal conductivity is shown in Fig.~\ref{fig:silicon}e.
\textcolor{black}{We compare the phonon properties obtained using quantum and classical statistics, while in Fig.~\ref{fig:silicon}e thermal conductivity calculations are reported also comparing the direct inversion of the scattering matrix with the RTA approach.}
The reference result is the calculation with quantum statistics and direct inversion of the scattering tensor, which gives $\kappa_{Q,Inv}=147$ \WmK, in good agreement with experiments\cite{Kremer:2004dv, Inyushkin:kw} and former BTE calculations.\cite{Jain:2015by,McGaughey:2019fa} As one should expect, solving BTE with RTA leads to underestimating the thermal conductivity, in this case by about 5$\%$, as we get $\kappa_{Q,RTA}=140$ \WmK. While this error for silicon at room temperature is acceptable, inaccuracies introduced by using RTA strongly depend on the phonon properties of the materials under study and become more severe at low temperature, where hydrodynamic effects are predominant.\cite{Cepellotti:2015ke} 

\begin{figure}[t!]
\includegraphics[width=0.95\linewidth]{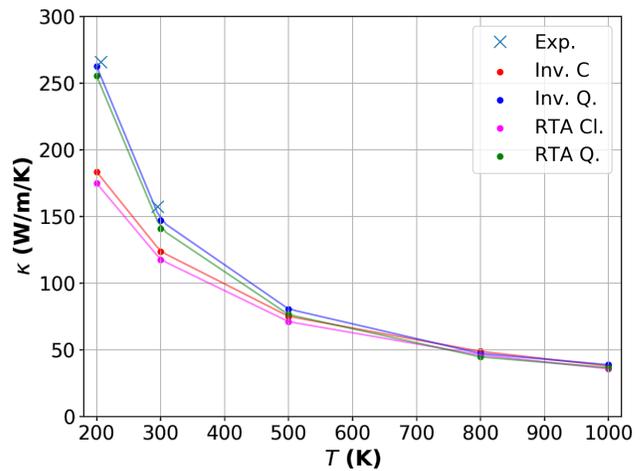}
\caption{
\textcolor{black}{Thermal conductivity of bulk Silicon as a function of  temperature, calculated with RTA and full inversion. Quantum with inversion (blue), quantum using RTA (green), classical with inversion (red) and classical using RTA (magenta) results are compared. As a reference, the experimental values are shown (light blue x).\cite{Inyushkin:kw}}}
\label{fig:silicon_vs_temp}
\centering 
\end{figure}

Using classical statistics, as discussed in section~\ref{sec:classical_limit}, leads to a reduction of the phonon lifetimes due to the increased population of the high-frequency modes. While in the classical limit every mode contributes to $\kappa$ with larger heat capacity, at 300~K this effect is overcome by the reduction of $\tau$, thus leading to an overall reduction of the thermal conductivity. The thermal conductivity of silicon at room temperature in the classical limit turns out $\kappa_{Cl,Inv}=$123 \WmK\ with direct inversion and $\kappa_{Cl,RTA}=$118 \WmK with RTA. \textcolor{black}{When increasing the temperature, the quantum and classical calculations converge, as illustrated in Fig.~\ref{fig:silicon_vs_temp}. 
}

As in the next section we discuss silicon clathrates treated with empirical potentials, it is worth noting that performing the same calculations using the Tersoff potential\cite{Tersoff:1989br} leads to a substantial overestimate of the thermal conductivity of diamond silicon with $\kappa_{Q}=$305~\WmK and $\kappa_{Cl}=260$~\WmK, as reported in previous works.\cite{He:2012hg}

\subsection{Pristine and Germanium-doped Type-II Silicon Clathrates}

\begin{figure*}
\centering
\includegraphics[width=1\linewidth]{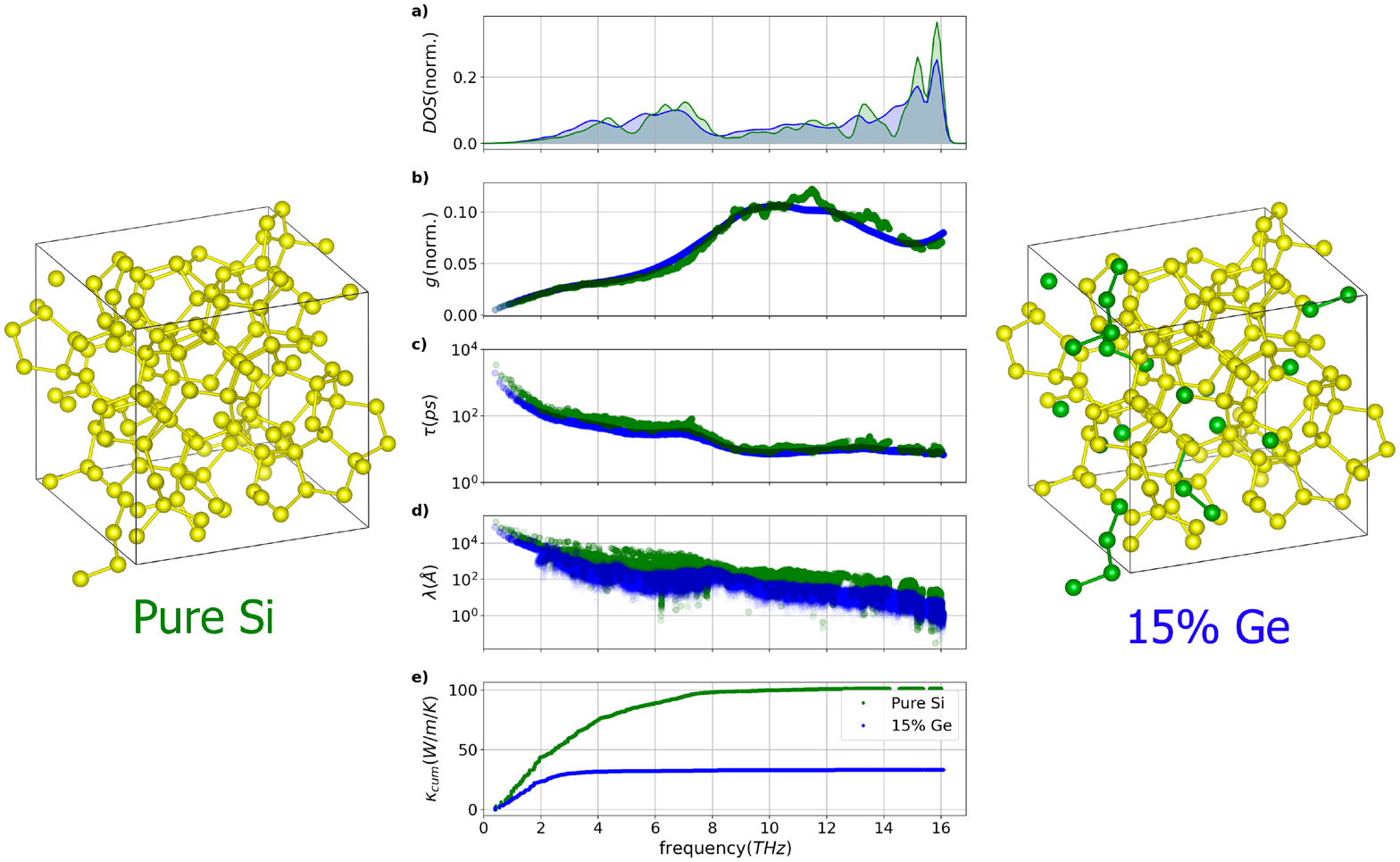}

\caption{\label{fig:clathrate_BTE}
Comparison between pristine type-II silicon clathrate Si$_{136}$ and the Si$_{116}$Ge$_{20}$ clathrate alloy carried out solving the BTE in the relaxation time approximation on a $136$ atoms cell. a) normalized density of states, b) normalized phase-space per mode $g$} c) lifetimes $\tau$ d) mean free paths $\lambda$ and e) cumulative thermal conductivity $\kappa_{cum}$. 
The values of $\tau$, $\lambda$, and $\kappa_{cum}$ are computed using quantum statistics for the phonon population at $300K$.
\end{figure*}

Silicon clathrates are low-density allotropes of silicon, consisting of a 3D network framework of covalent tetrahedral bonds.  These networks form cages that can incorporate extrinsic guest atoms, such as alkaline and alkaline-earth atoms.\cite{Kasper1713}
Because of these characteristics, clathrates are being actively studied for batteries, and hydrogen storage.\cite{ShojiYamanaka:2014bj, XihongPeng:2015jx} 
These structures \textcolor{black}{embody} the electron-crystal-phonon-glass concept, which has been proposed as a paradigm for efficient thermoelectric materials, with low thermal conductivity and high power factor.\cite{nolas2002figure, Beekman:2015cy, Beekman:2016ff}

Among silicon clathrates, two structures are commonly synthesized: type-I and type-II with 46 and 136 Si atoms per conventional cubic cell, respectively. 
The irreducible cells for these structures contain 23 and 34 atoms, which made it possible to compute their thermal conductivity by {\sl ab initio} ALD.\cite{harkonen2016ab, norouzzadeh2017thermal} These works found that the thermal conductivity at room temperature of both clathrates is about 50 \WmK, that is $\sim$1/3 that of bulk silicon. Further thermal conductivity reduction may be achieved introducing substitutional defects in the cage structure and rattlers.\cite{He:2014ge, Chen:2018bt} 
The intercalation with rattling cations\cite{Chen:2018bt} alters the dispersion relations and the scattering rates of clathrates phonon modes and their thermal conductivity. Without introducing guest atoms in the cages, heat transport can be engineered by  substituting silicon in the framework with isovalent atoms, e.g. germanium, that maintain charge neutrality.\cite{Shen:2020jx}

Applying the theoretical framework outlined in section~\ref{sec:methods} we investigate the effect of substitutional germanium on the thermal conductivity of the type-II silicon clathrate, transitioning from ordered to random alloy configurations. 
%
\textcolor{black}{The following investigation is performed} at a temperature of 300~K and it develops in three parts. 
First, since \appname\ allows to carry out BTE calculations for large systems, we consider the 136 atoms unit cell, composed of 4 face sharing dodecahedral (Si$_{20}$) and 2 hexakaidecahedral (Si$_{28}$) silicon cages, and then we analyze the effects of substituting 20 Germanium atoms ($15\%$) in this structure. 
Second, we replicate this structure in a $5\times 5\times 5$ supercell ($17000$ atoms) and we verify that $\kappa$ obtained by QHGK complies with the BTE calculation. 
Third, we use QHGK to compute thermal transport in a random alloy clathrate which has a much lower thermal conductivity than the corresponding alloy with diamond structure. 

\begin{figure*}[ht]
\centering
\includegraphics[width=1\linewidth]{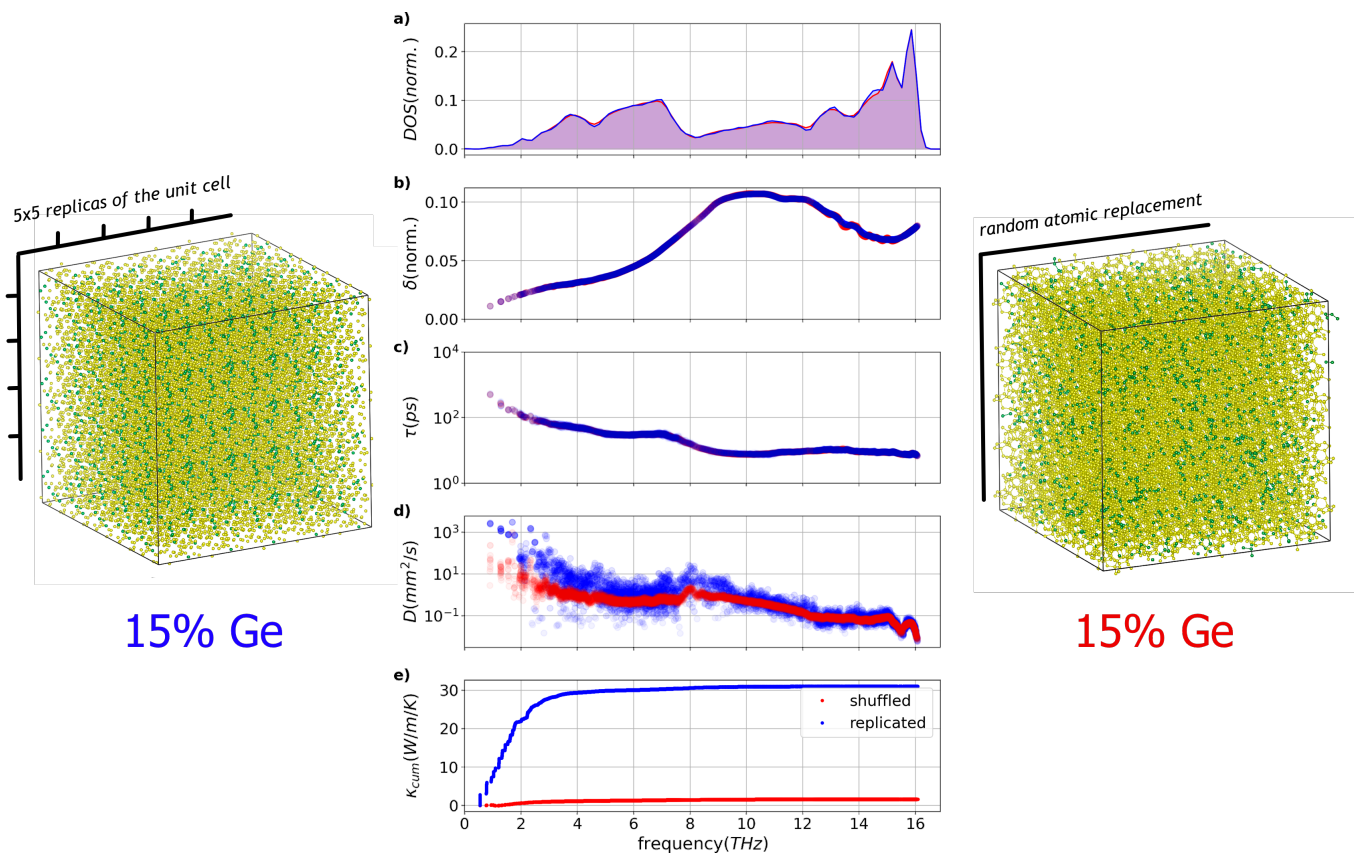}
\caption{\label{fig:clathrate_QHGK}
Comparison Type-II Silicon clathrate with and Germanium concentration at $15\%$ using QHGK on a $17000$-atoms unit cell. On the left, the Germanium atoms have been substituted in the unit cell, and then the unit cell has been replicated. On the right, the unit cell has been replicated and then the Silicon has been replaced by the Germanium atoms.
a) density of states, b) normalized phase-space per mode $g$ c) lifetime $\tau$, d) diffusivity per mode $D$, and e) cumulative thermal conductivity.
The values of $\tau$, $\lambda$, and $\kappa_{cum}$ are computed using quantum statistics for the phonon population at $300K$.}
\end{figure*}

In the BTE calculations, starting from a minimized structure with lattice parameter $14.62$~\AA, the second and third IFCs in Eq.~\ref{eq:second_and_third_ifc} are computed using the Tersoff empirical potential,\cite{Tersoff:1989br} the ASE package\cite{ase-paper} and the unfolding technique described in Appendix~\ref{sec:implementation_details}, on the 136-atoms conventional cell.
As the Tersoff potential is short-range ($r_{cut}$=3.1~\AA) the second and the third derivatives of the potential are calculated for the full range of the interatomic interactions by finite differences. 
The ALD calculations are performed with \appname\ using a (7, 7, 7) q-point grid, which is sufficiently dense to obtain well converged thermal conductivity. To solve the BTE we use the RTA solver (Equation~\ref{eq:RTA}). Corrections beyond RTA (Eq.~\ref{eq:full_conductivity}), amount to a less than 1\% increase of the thermal conductivity.\cite{harkonen2016ab}

Fig.~\ref{fig:clathrate_BTE} compares the phonon transport properties of the pristine silicon clathrate (Si$_{136}$) and the 15\% Germanium substituted one (Si$_{116}$Ge$_{20}$), in which substitution sites were chosen randomly.  
The room temperature thermal conductivity of Si$_{136}$ is 101.3~\WmK, which is about 1/3 of the reference value for diamond silicon computed by BTE with the Tersoff potential ($\kappa=305$ \WmK).\cite{He:2012hg} This result is in agreement with former first-principles BTE studies that report similar thermal conductivity ratios between diamond silicon and clathrates.\cite{harkonen2016ab, norouzzadeh2017thermal}
In the 15\% Ge-alloyed material the thermal conductivity drops to 33.0~\WmK, which is a further \textcolor{black}{$3\times$} factor. 
The short-range disorder introduced by the Ge atoms breaks the symmetry of the system, modifying phonon dispersion relations and density of states. While the impact on the scattering phase-space volume appears small (Fig.~\ref{fig:clathrate_BTE}b), phonon lifetimes (Fig.~\ref{fig:clathrate_BTE}c) are reduced across the whole spectrum of frequencies. Group velocities are also reduced especially at intermediate frequencies, thus leading to much shorter phonon mean free paths, in particular in the frequency range between 1 and 8 THz (Fig.~\ref{fig:clathrate_BTE}d). The effect on the frequency-resolved cumulative thermal conductivity is shown in Fig.~\ref{fig:clathrate_BTE}e.

\begin{table}[h]
\begin{center}
\begin{tabular}{ c  c  c  }
\hline
Structure & Method & $\kappa$ \WmK \\
\hline
\hline
 Si$_{136}$ & \emph{BTE} & 101.3 \\ 
\hline
Si$_{116}$Ge$_{20}$  & \emph{BTE} & 33.0 \\
\hline
Si$_{116}$Ge$_{20}$  & \emph{QHGK} &  \textcolor{black}{31.0}\\
\hline
Si$_{0.85}$Ge$_{0.15}$ clathrate  & \emph{QHGK} & \textcolor{black}{1.7} \\
\hline
\end{tabular}
\end{center}
\caption{\label{tab:conductivities}
Comparison of thermal conductivities calculated from different Silicon-Germanium alloys using the BTE and QHGK. The discrepancy between the $15\%$-concentration replicated clathrates calculated using QHGK and BTE is due to size scale effect.\cite{Isaeva:2019dm}}
\end{table}

While introducing Ge breaks the symmetry within the conventional cubic cell of Si$_{136}$, the system studied so far retains long-range crystalline order. To study a random alloy one would need either to introduce an effective mass scattering term in the BTE\cite{Tamura:1983vp,Garg:2011hi} or to randomly substitute Ge in a replicated supercell. 
The supercell approach rapidly exceeds the reach of our BTE implementation when the system is larger than $\sim 1000$ atoms, but it is tractable by QHGK. 
QHGK allows us to compare thermal transport in pseudo-random clathrate alloys for 15$\%$ Ge substitution in the 136-atom conventional cell and in a $5\times 5\times 5$ replicated supercell (17000 atoms), the latter being a more realistic representation of an actual random alloy clathrate. The two structures and their phonon transport properties are shown in Figure~\ref{fig:clathrate_QHGK}, in which the 136-atom system is also replicated $5\times 5\times 5$.
The results are summarized in Table~\ref{tab:conductivities}.
The thermal conductivity of the 136-atom system computed by QHGK on a $5\times 5\times 5$ replicated supercell at the $\Gamma$ point is  \textcolor{black}{31.0~\WmK, slightly lower than the reference BTE calculation, proving the agreement between BTE and QHGK and indicating that the 17000-atom system is close to size convergence. 
While all the properties of this system are calculated for the $5\times 5\times 5$ supercell, we interpolate the phonon lifetimes as a function of frequency from the explicit ALD calculation of a $3\times 3\times 3$ system (3672 atoms) using third-order splines.\cite{Isaeva:2019dm}}
\textcolor{black}{
The effect of randomly substituting Ge atoms in the 17000-atom supercell on the thermal conductivity is striking, as $\kappa$ drops to 1.7~\WmK.} 
The overall thermal conductivity reduction with respect to bulk (diamond) silicon is an impressive $200\times$ factor. 
Whereas phonon DOS and lifetimes are not affected by long-range disorder (Fig.~\ref{fig:clathrate_QHGK}a-c) What actually makes a large difference is the phonon diffusivity $D(\omega)$ (Fig.~\ref{fig:clathrate_QHGK}d), defined in Eq.~\ref{eq:diffusivity_per_mode}. 
The latter is a function of the generalized phonon velocities defined in Eq.~\ref{eq:generalized_velocity}. This quantity is related to the delocalization of each mode and to its ability of transferring energy to modes with similar frequency thus enabling thermal transport in non-periodic materials. 
Fig.~\ref{fig:clathrate_QHGK}d-e shows that low-frequency long-wavelength modes are the most affected by the increased range of disorder, and the difference in thermal conductivity between the two systems builds up for frequencies lower than 3~THz.

\section{Conclusions}
In summary, we reviewed the background theory and a new implementation of ALD approaches to compute the thermal conductivity of solids. The widely used BTE approach is supplemented by a recently developed unified theory, QHGK, that extends the application of ALD to compute heat transport in disordered materials.
Using a relatively complex benchmark system, Ge-alloyed type-II silicon clathrate, we have shown that QHGK gives results in agreement with BTE for partially disordered systems, and it allows one to compute the thermal conductivity of systems with longer-range disordered. Through this unified approach, we analyzed the role of different levels of crystal symmetry breaking in the clathrate structure, showing that by alloying with Ge the Si$_{136}$ clathrate structure a stark reduction of the thermal conductivity can be attained, about a factor 200 lower than that of crystalline diamond silicon. 

The methodology illustrated in this article is implemented in \appname, a modern Python-based software that implements both the BTE and the QHGK methods, and can seamlessly link to different molecular simulation packages to get IFC either from DFT or semiempirical forcefields. \appname\ runs on GPUs and CPUs, and is released open-source, for the scientific community to use and develop further.

\section*{Data Availability}
The \appname\ package is available for download at \textcolor{blue}{\href{https://github.com/nanotheorygroup/kaldo}{https://github.com/nanotheorygroup/kaldo}}. Input files to produce the data that supports the findings of this study are available in the same repository in the ``examples" folder.

\begin{acknowledgments}
The authors are grateful to Riccardo Dettori, Charles Sievers, Shunda Chen, and Alfredo Fiorentino for many insightful suggestions. 
G. B. particularly thanks Doaa Altarawy for valuable discussions and code reviews. 
G. B. gratefully acknowledges support by the Investment Software Fellowships (grant No. OAC-1547580-479590) of the NSF Molecular Sciences Software Institute (MolSSI) (grant No. OAC-1547580) at Virginia Tech.\cite{molssi1, molssi2}
\end{acknowledgments}

\appendix

\section{Code details}\label{sec:implementation_details}

\begin{figure}
\centering
\includegraphics[width=0.95\linewidth]{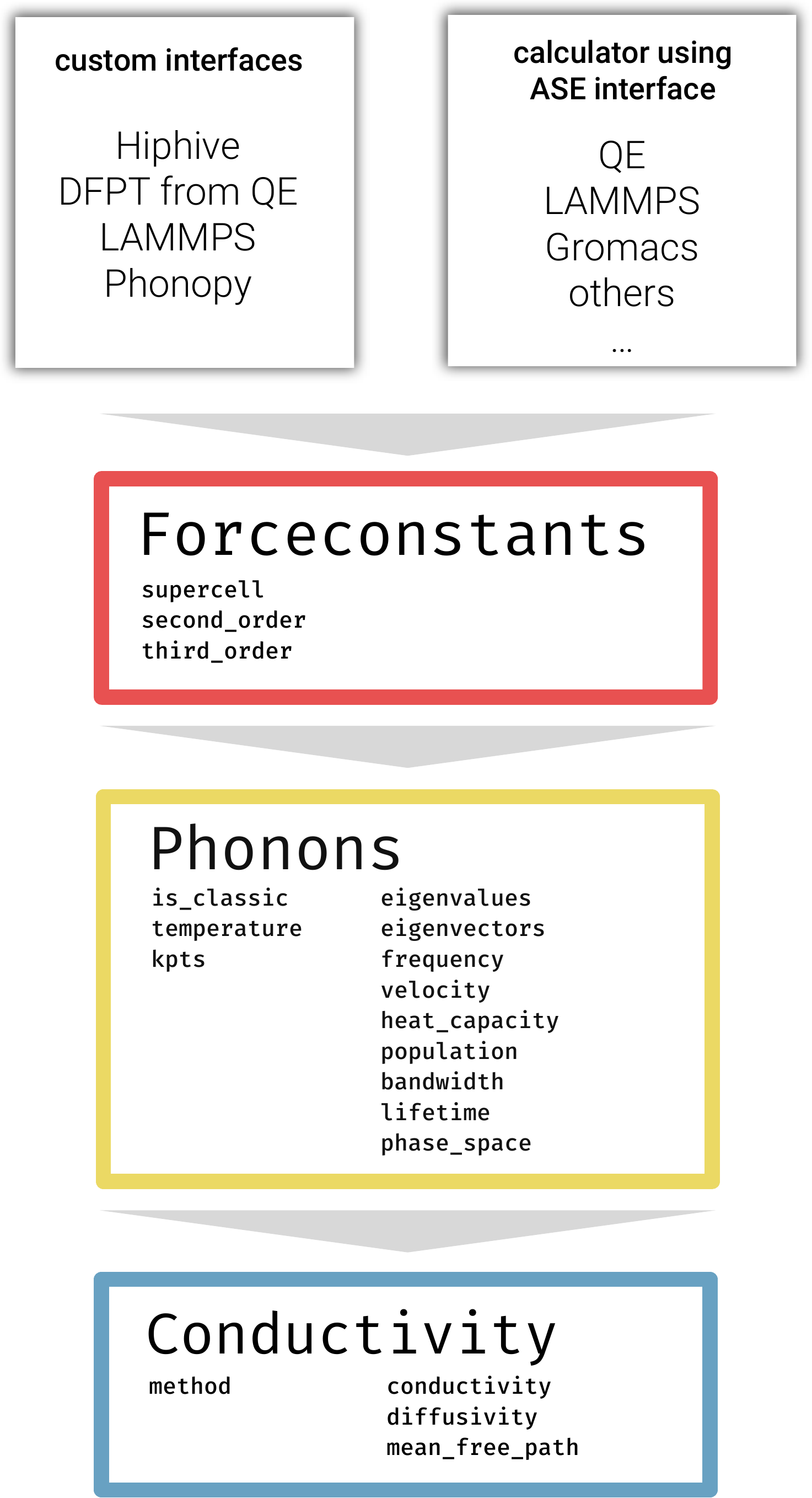}
\caption{\label{fig:architecture}
The modular software architecture of \appname.}
\end{figure}

The main features of \appname, have been highlighted in the main text. We here want to point out some technical details that can be useful in performing ALD calculations and to describe the simulation workflows. 
The software architecture is illustrated in Fig.~\ref{fig:architecture}. 
The input needed to create the IFC can be provided in two different ways, using an ASE calculator, which provides a general connector to many quantum chemistry codes or using one of the custom interfaces. When using ASE, the IFC calculated using finite displacements, and it depends on the choice of the shift, which is specified in $A$, and on the choice of the supercell. 

For large unit cells, \appname\ is able to exploits the periodic boundary conditions to reduce the number of calculations. If the interatomic forcefield is negligible outside a distance threshold and the unit cell is larger than twice this threshold, we can calculate forces only inside the unit cell, and exploit the periodic boundary condition, replacing the second IFC $\phi^{\prime\prime}_{il'i'}\rightarrow \phi^{(2)}_{i0i'} $  and the third $\phi^{(3)}_{il'i'l''i''}\rightarrow\phi^{(3)}_{i0i'0i''}$, effectively reducing the dimension of the system from $N_{replicas}\cdot N_{atoms}\to N_{atoms}$.  The dynamical matrix is stored as a dense tensor of size $(3 N_{atoms})^2$, or $N_{replicas}(3 N_{unit})^2$ when using a supercell. The third order IFC is a sparse tensor of size $(3 N_{atoms})^3$, or $(N_{replicas})^2(3 N_{unit})^3$.

The phonons calculations are dependent on the temperature,  on the choice of the statistics, and on the choice of the size of the mesh of the reciprocal space. 
An important remark needs to be made about the Dirac delta in Eq.~\ref{eq:gamma_operators}. The first delta in Eq. ~\ref{eq:delta_deconstructed} corresponds to the quasimomentum conservation in three-phonons scattering processes, in crystal. This symmetry can be exploited directly, as a constraint, to reduce the memory needed to store the single phonon state from $(N_k (3 N_{unit}))^2$ to  $2(N_k) (3 N_{unit})^2$, and the factor $2$ comes to the two possible processes of creation and annihilation.
The second $\delta$-distribution in Eq.~\ref{eq:delta_deconstructed} guarantees the energy conservation in all scattering processes. It requires a numerical implementation of the Dirac delta. Three common implementations are provided with the code: Gaussian, Lorentzian, and triangular. When introducing a frequency threshold, a finite-width representation of the Dirac delta can be exploited to store the generalized velocity as a sparse tensor. For a sufficiently dense energy grid, the physics is not dependent on the choice of the value for $\delta$.\cite{Fugallo:jl} In crystals, for a given grid size, an optimal choice of the width of the $\delta$-function can be estimated from the phonon velocities.\cite{Li:2014fc, Li:2012bd}

In the calculation of the conductivity, we need to specify the engine to use for the calculations, among "inverse", "self-consistent", and "RTA" for crystals, and "QHGK" for amorphous systems. In the latter, we can also provide an optional bandwidth. In this case the QHGK calculations are equivalent\cite{Isaeva:2019dm} to the Allen and Feldman model.\cite{Allen:1993gt} When performing QHGK calculations, and scaling to large systems, it is possible to introduce an energy threshold on the Lorentzian $\tau_{\mu\mu'}$ in Eq.~\ref{eq:lorentzian_amorphous}. Such tensor has size $(3N_{atoms})^2$, and is sparse when we introduce a threshold such that $\tau_{mn}=0$ when $|\omega_{n}-\omega_{m}|<\omega_{threshold}$.

\bibliography{bibliography.bib}

\end{document}